\documentclass{aa} 
\usepackage{graphicx,natbib} 
\usepackage{amssymb} 
\newcommand{\mybf}[1]{#1} 
\newcommand{\mymbf}[1]{\ensuremath{#1}} 

\newif\ifpdf 
\ifx\pdfoutput\undefined 
\pdffalse % we are not running pdflatex 
\else 
\pdfoutput=1 % we are running pdflatex 
\pdftrue 
\fi
 
\def\aa_{\aap} 
\def\pfl{Phys.\ Fluids} 
\def\plb{Phys.\ Lett.\ B} 
 
\def\rmp{Rev.\ Mod.\ Phys.}

\def\mas{MMS67} 
\def\much{M84} 
\def\iben{IM85} 
\def\paq{P86} 
\def\burg{B69} 
\def\hahn{H71} 
\def\thoul{TBL} 

\bibpunct{(}{)}{;}{a}{}{,} % to follow aa-style 

\newcommand{\citeposs}[1]{\citeauthor{#1}'s \citeyearpar{#1}} 
 
\newcommand{\bref}[1]{\mbox{(\ref{#1})}} 
\begin{document}

\title{Quantum corrections to microscopic diffusion constants}  
 
\author{H.~Schlattl and M.~Salaris} 
 
\institute{ 
Astrophysics Research Institute, Liverpool John Moores 
University, Twelve Quays House, Egerton Wharf, Birkenhead CH41 1LD, 
United Kingdom} 
 
\offprints{Helmut Schlattl, \\ \email{hs@astro.livjm.ac.uk}} 
 
\date{Received / Accepted} 
 
\titlerunning{Quantum corrections to diffusion constants} 
 
\abstract{We review the state of the art regarding the 
computation of the resistance coefficients in conditions typical of 
the stellar plasma, and compare the various results studying their 
effect on the solar model.  
We introduce and discuss for the first time in \mybf{an} astrophysical
\mybf{context} 
the effect of quantum corrections to the evaluation of 
the resistance coefficients, and 
provide simple yet accurate fitting formulae 
%resistance coefficients,
%inclusive of quantum effects.
\mybf{for their computation. Although the inclusion of quantum corrections
only weakly modifies the solar model, their effect is growing with
density, and thus might be of relevance in case of denser
objects like, e.g., 
white dwarfs.} 
} 
 
\maketitle 
 
\section{Introduction}  
 
Basic physical considerations suggest that, in addition to convective 
mixing -- routinely included in stellar evolution computations -- 
additional transport processes are efficient within the stellar 
interior; they are driven by pressure,  
temperature and chemical abundance gradients, and by the effect of radiative 
pressure on the individual ions. 
These processes are collectively called 'diffusion processes', and 
their inclusion in stellar evolution computations is necessary in 
order to satisfy the helioseismological constraint for the solar 
models \citep{BahPin}. 
 
In general, individual ions are forced to move under the influence of 
pressure as well as temperature gradients, which both  
tend to move the heavier elements toward the centre of 
the star, and of concentration gradients that oppose the above processes. 
Radiation, which causes negligible diffusion in the Sun,
pushes the ions toward the surface, whenever the radiative 
acceleration of an individual ion 
species is larger than the gravitational acceleration. 
The speed of the diffusive flow depends on 
the collisions with the surrounding 
particles, as they share the acquired momentum in a random way. 
It is the extent of these 'collision' effects that dictates the 
timescale of element diffusion within the stellar structure, once the  
physical and chemical profiles are specified. 
The most general treatment for the element transport in a 
multicomponent fluid associated with 
diffusion is provided by \citeauthor{Burgers}' (\citeyear{Burgers}, \burg)
equations.  
In these equations the effect of collisions between ions  
is represented by the so-called resistance coefficients, i.e. 
the matrices $K$, $z$, $z^\prime$, 
$z^{\prime\prime}$, whose precise evaluation is fundamental 
in order to estimate correctly the diffusion timescales for the 
various elements\footnote{In the original \burg\ book the elements  
of the matrix $K$ are called resistance coefficients,  
whereas $z$, $z^\prime$, $z^{\prime\prime}$, 
which are related to the efficiency of the thermal diffusion, have no 
specific names. In this paper we follow the notation of \cite{PPF86}, 
who termed both $K$ and $z$, $z^\prime$ and $z^{\prime\prime}$ as 
resistance coefficients.}. 
 
Recent spectroscopic determinations of Fe and Li abundances in
Galactic globular  
cluster turn-off stars discussed in, e.g., \citet{GBB01} or \citet{BPS02}, 
have shown that 
the present standard treatment of diffusion is in disagreement 
with the observed surface abundance of these two elements in 
stars of globular clusters \mybf{(for the Li problem see, e.g.,
\citealt{MFB84} or \citealt{VC95})}. In the light of these results, it
is very important to conclusively assess how much of this discrepancy
is due to an incorrect treatment  
of the diffusion process, and how much is due to competing 
rotationally induced \mybf{or other non-standard macroscopic} mixing phenomena, which inhibit the  
efficiency of diffusion. 
 
In this paper the state of the art regarding the 
computation of the resistance coefficients in conditions typical of 
the stellar plasma is reviewed and a detailed 
comparison of the results from different authors is performed. The 
effect of different resistance coefficients on solar models is
examined, too. Moreover,   
we introduce and discuss for the first time in astrophysical computations
the effect of quantum corrections to the evaluation of 
the resistance coefficients. 
We also provide simple fitting formulae for accurate calculations
of resistance coefficients including the appropriate quantum corrections.  
In \S~2 we compare existing determinations of the resistance coefficients, 
and discuss the differences on solar models; in 
\S~3 we determine the appropriate quantum corrections to 
these coefficients, and conclusions will follow in \S~4. Analytical formulae for the 
computations of updated resistance coefficients, 
including quantum corrections, are given in the
appendices\footnote{FORTRAN77 routines to compute the improved resistance
coefficients (including quantum corrections) and their implementation
into \citeposs{S02} diffusion-constant 
routine are publicly accessible under
\texttt{http://www.astro.livjm.ac.uk/$\mathtt{\tilde{~}}$hs}.}.
 
\section{Classical treatment of element diffusion} \label{coll} 
 
The diffusion of elements in a multicomponent fluid can be treated 
either according to the \citet{CC70} or the 
\burg\ formalism. We are following 
%in which case the
%treatment by \burg\ is more convenient. For a 
%collision-dominated plasma like the stellar one, 
%the two formalisms are equivalent, and the 
%transport properties computed with the two methods must be the same. 
%Throughout the rest of the paper we will be following  
the latter description, which is equivalent to the so-called ``second 
approximation'' in \citet{CC70}. It would be desirable to use a higher
order approximation, because an uncertainty of the order of 10\% is
introduced by using only the second one~\citep{RD82}, independent
of the accuracy in the resistance coefficients. But this can only be done
in the scheme of \citet{CC70}, which becomes very cumbersome for
gases with more
than two components. 
%should in principle provide better accuracy, 
%but terms beyond the second one introduce only small corrections  
%\citep[\paq]{PPF86}.  
 
\subsection{Existing calculations of resistance coefficients} 
 
Burgers' equations are obtained assuming  
the gas particles have approximate Maxwellian velocity distributions; 
the temperatures are the same for all particle species; the mean 
thermal velocities are much larger than the diffusion velocities; 
magnetic fields are unimportant. 
Burgers' scheme involves the  
resistance coefficients $K$, $z$, $z^\prime$  
and $z^{\prime\prime}$; following \citet[\mas]{MMS67},   
they can be expressed in terms of the so-called 
reduced collision integrals 
${\Omega^{(l,s)}_{ij}}^\ast$~\citep{HCB54}\footnote{Note the error in the 
table at  
page 44 of \burg: \citeposs{HCB54} $8 \mu 
\Omega^{(ij)}_{st}$ is equivalent to  
$\Sigma_{st}^{(ij)}$ in \burg\ and \citeposs{CC70}  
$8 \mu \Omega^{(i)}_{st}(j)$.}, 
according to the following relationships: 
\begin{eqnarray} 
\frac{K_{ij}}{K_{ij}^0} & = & 4 
\frac{{T^\ast_{ij}}^2\,{\Omega^{(1,1)}_{ij}}^\ast}{\ln\left(\Lambda_{ij}^2 
+1\right)}\mathrm{,} \label{kdef} \\ 
z_{ij} & = & 1 - 1.2 
\frac{{\Omega^{(1,2)}_{ij}}^\ast}{{\Omega^{(1,1)}_{ij}}^\ast}\mathrm{,} \\ 
z_{ij}^\prime & = & 2.5 - \frac{6\,{\Omega^{(1,2)}_{ij}}^\ast - 4.8\, 
{\Omega^{(1,3)}_{ij}}^\ast}{{\Omega^{(1,1)}_{ij}}^\ast}\mbox{\quad and} \\ 
z_{ij}^{\prime\prime} & = & 2 
\frac{{\Omega^{(2,2)}_{ij}}^\ast}{{\Omega^{(1,1)}_{ij}}^\ast}\mathrm{,} \label{zppdef} 
\end{eqnarray} 
where 
\begin{eqnarray} 
T^\ast_{ij} & = & k_\mathrm{B}T\frac{\lambda_\mathrm{D}}{\left|Z_i Z_j 
e^2\right|}\quad\mbox{and}\label{Tdef} \\ 
K_{ij}^0 & = & \frac{2}{3}\sqrt{\frac{2\mu_{ij} \pi}{(k_\mathrm{B} T)^3}} 
\left(Z_i Z_j e^2\right)^2 n_i n_j \ln\left(\Lambda_{ij}^2 +1\right)  
\end{eqnarray} 
with $\lambda_\mathrm{D} = \sqrt{k_\mathrm{B}T/(4\pi 
n_\mathrm{e} e^2)}$ being the Debye-length, $\Lambda_{ij} = 
4 T_{ij}^\ast$ the plasma parameter, $\mu_{ij} = m_i 
m_j/(m_i+m_j)$ the reduced mass, and $Z_i$, $m_i$ and $n_i$ the 
charge number, mass and particle number density 
of species $i$, respectively.  
%Here, we adopted basically the formalism of 
%\citet[\mas]{MMS67}.  
The collisions between the particles of the stellar plasma  
determine the values of the ${\Omega^{(l,s)}_{ij}}^\ast$ 
integrals; the physics of the collisions is specified by some form of 
the Coulomb interaction which, as a first approximation,
can be described by a pure Coulomb potential with a long-range cut-off  
distance, usually assumed to be $\lambda_\mathrm{D}$. 
According to \burg, using this truncated pure Coulomb potential the 
resistance coefficients become  
\begin{eqnarray} 
\frac{K_{ij}}{K_{ij}^0} & = & 2\, \frac{\ln\Lambda_{ij} - C_\mathrm{E} 
\pm \frac{\pi^2}{4}}{\ln\left(\Lambda_{ij}^2 + 1\right)} \label{Kburg}{\rm,} \\  
z_{ij} & = & 0.6{\rm,} \label{zb} \\ 
z^\prime_{ij}  & = & 1.3\mbox{\quad and} \label{zpb}\\ 
z^{\prime\prime}_{ij} & \approx & 2\mathrm{,} \label{zppb} 
\end{eqnarray} 
where $C_\mathrm{E}$ is Euler's constant and the alternative signs in 
Eq.~\bref{Kburg} represent  
repulsive ($+$) and attractive ($-$) particles, respectively. For 
high densities and thus small values of the plasma parameter, the 
resistance coefficients become negative (for an attractive potential 
already for $\Lambda \lesssim 20$). Since in stellar plasmas, in 
particular in case of white dwarfs, $\Lambda$ is often much smaller than this 
value, more elaborate computations have been performed.  

\begin{table}[t] 
\caption{Quantities in this work, which follows basically \mas, 
and their relation to quantities used by other authors.\label{tab:equi}} 
\begin{tabular}{cccc}\hline \hline 
This work & \much & \iben & \paq \\ \hline 
$\Lambda_{ij}$ & $\Lambda_{ij}$ & $x_{ij}$ & $\gamma_{ij}$ \\ 
${T_{ij}^\ast}^2 {\Omega^{(s,t)}_{ij}}^\ast$ &  --- & --- & 
$\frac{1}{4}F_{ij}^{(st)}$ \\  
$\Phi_{ij}$ & $\Phi$ & $\Phi$ & $\Psi_{ij}$ \\ 
$K^0_{ij}$ & $(K_{ij})_0$ & $K^0_{ij}$ & $\epsilon_{ij}\times\frac{16}{3} x_i x_j n 
m M_i M_j  \mathrm{e}^{\Psi_{ij}}$ \\ \hline 
\end{tabular} 
\end{table} 

More accurate collision integrals have been obtained by considering 
a Debye-H\"uckel type of potential 
\begin{equation} 
V_{ij}(r) =  \frac{Z_i Z_j e^2}{r}\; \mathrm{e}^{-r/\lambda_\mathrm{D}} \label{debhuck} 
\end{equation} 
where $r$ is the particle distance. 
The results are provided either in tabulated form by \mas\ 
or as fitting formulae by \citet[\much]{Much84}, \citet[\iben]{IM85} 
and \citet[\paq]{PPF86}. In the latter cases the integrals are given as 
functions of   
\begin{equation}\label{phidef} 
\Phi_{ij} = \ln\left(\ln(1+\Lambda_{ij}^2)\right) \label{fi}. 
\end{equation} 
Different notations for the same physical 
quantities have been used by these groups; in Table~\ref{tab:equi} 
their relations to the present work are summarized.  
\much, \iben\ and \paq\ argue that while $\lambda_\mathrm{D}$ is an 
appropriate screening distance at low densities, the Debye sphere 
loses its significance at high densities, in which case a more 
appropriate screening distance is the mean interionic distance; thus 
they suggest to use in the actual stellar model computations the 
larger of $\lambda_\mathrm{D}$ or the mean interionic distance. 
Whatever one chooses for the actual screening distance, its
value has to be employed as $\lambda_\mathrm{D}$ in Eq.~\bref{Tdef} to
obtain the appropriate value for $\Lambda_{ij}$ for determining the collision
integrals. We also remark that, at 
least for the entire evolution of the Sun up to now,   
$\lambda_\mathrm{D}$ has always been larger than the mean  
interionic distance. 

\begin{figure}[t] 
\includegraphics[width=\columnwidth]{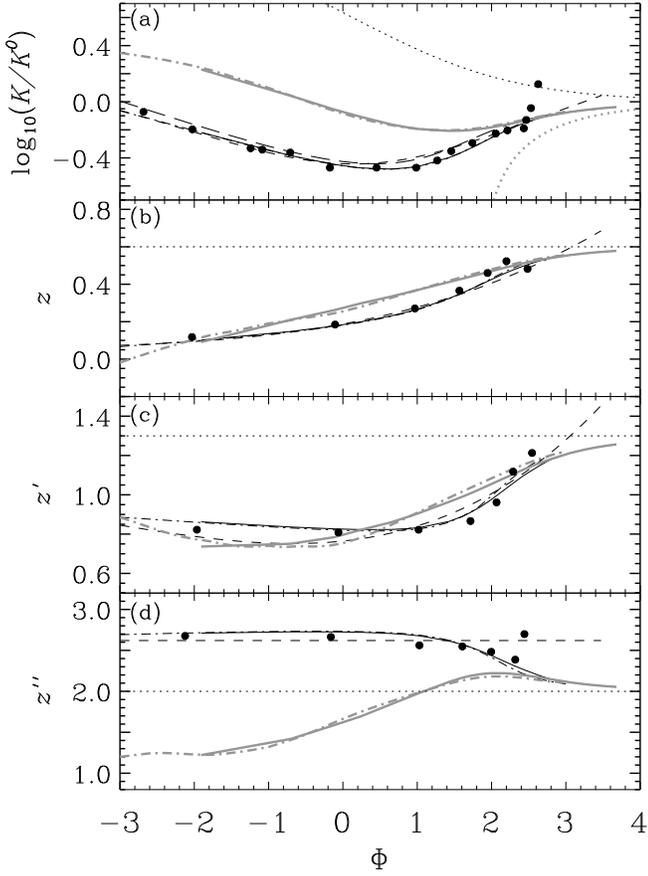} 
\caption{The resistance coefficient $K$ (a), $z$ (b), $z^\prime$ (c) and 
$z^{\prime\prime}$ (d) as computed by \burg\ 
(dotted), \mas\ (solid), \much\ (short-dashed), \paq\ (dash-dotted) 
and \iben\ (long-dashed line). The dark black lines indicate the case 
of repulsive potentials, while the brighter ones show the values for 
attractive forces (not computed by \much\ and \iben). The dots represent the 
actual values computed by  
\much.\label{fricvgl}} 
\end{figure} 

The results of the different groups are compared in 
Fig.~\ref{fricvgl}.  
Noticeably, given that they use 
all the same physical assumptions, they all got very similar 
values, but they all differ  
significantly from Burgers' results at higher densities (i.e., lower
values of $\Phi$), where his approximations are not
adequate anymore. 
\much's formulae for the case of a repulsive potential 
have been obtained by fitting solely the values represented by dots 
in Fig.~\ref{fricvgl}. Although the formulae work pretty
well overall, $z^{\prime\prime}$  
deviates considerably at $\Phi=2$ from the mean value of $2.6$, and 
the almost linear behaviour of $z^\prime$ for $-3<\Phi<1$ is only
poorly reproduced.  

\begin{figure}[t] 
\includegraphics[width=\columnwidth]{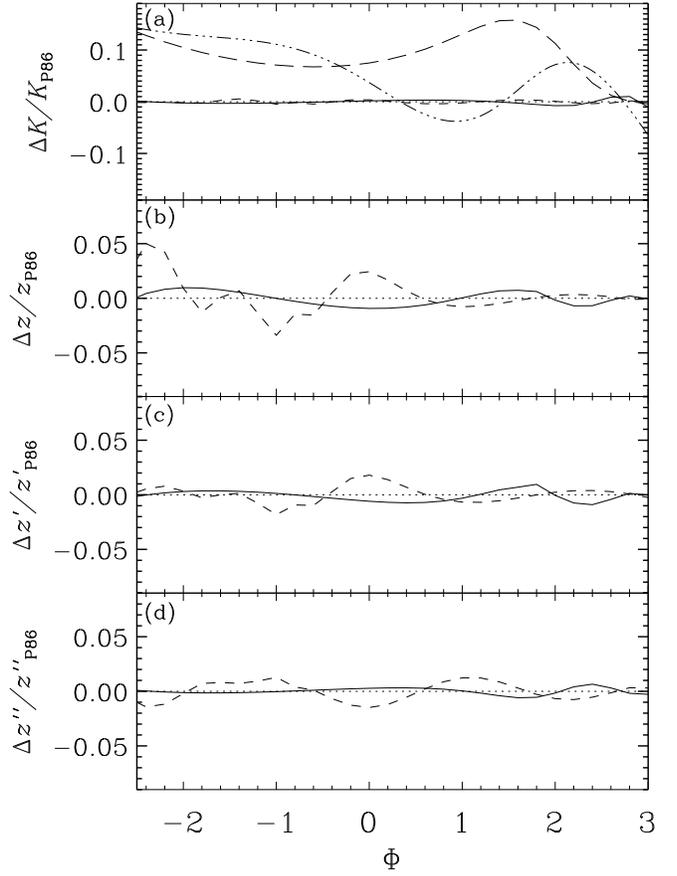} 
\caption{The relative difference of $K$ (a), $z$ (b), $z^\prime$ (c) and 
$z^{\prime\prime}$ (d) between our analytic fits and the 
values calculated by \paq. The solid line represents the case of a 
repulsive potential, the short-dashed one the case of an attractive 
potential; the dotted line indicates the line of zero difference.  
Also shown in (a) is the difference between the \iben's and \paq's results 
(long-dashed) and \iben's and \much's results (dash-dot-dot-dotted  
line).  
\label{fitvgl}} 
\end{figure} 

The supposedly most accurate calculation of the collision integrals 
has been performed in \paq, where cubic splines for 50 equally spaced 
intervals in $\Phi$ are provided; 
the results agree very well with the tabulation by \mas. 
To obtain more manageable but still accurate formulae,  
%which later will be use in conjunction 
%with the quantum corrections to the resistance coefficients,   
we fitted polynomials of at most 5$^\mathrm{th}$ 
order to their values for $-2.5<\Phi<3$; this enables to 
compute diffusion of elements up to Fe in stars on the main sequence 
and to follow with sufficient accuracy the early white dwarf cooling phase. 
The numerical values for the polynomial coefficients are given in  
Appendix A.  

In Fig.~\ref{fitvgl} the relative differences between our 
fitting formulae and the values of \paq\ are shown. For the case of a 
repulsive potential ($e$-$e$ or ion-ion collisions) the results of 
\paq\ for all quantities can be reproduced with an accuracy better 
than 2\%; for an 
attractive potential ($e$-ion collisions) the accuracy is  
still well within 5\%, and for a large range of 
$\Phi$ values it is certainly better than 3\%. Taking into 
account that the results of different groups for the same physical 
assumptions, e.g., between \iben\ and 
\paq\ (long-dashed line in Fig.~\ref{fitvgl}a), differ by up to 15\%, 
we consider our polynomial fits to be sufficiently accurate.  
 
In main sequence stars the value of $\Phi$ does not deviate 
considerably from $2$ for H and He (see Fig.~\ref{diffvgl}b), 
thus constant values for  $z$, $z^\prime$ and 
$z^{\prime\prime}$ are usually assumed. Moreover, in stellar 
plasmas the amount of element diffusion is basically determined by 
ion-ion collisions, and as  
the differences for attractive and 
repulsive potential are small when  $\Phi\sim 2$,  
 the resistance coefficients for a repulsive potential 
are often employed for all cases.  
For instance, the widely used diffusion routine by  
\citep[\thoul]{diffc} uses the values of \burg\ for $z$, $z^\prime$ and 
$z^{\prime\prime}$ (Eq.~\ref{zb}--\ref{zppb}) while for $K$ the 
fitting formula of \iben\ has been implemented. This assumption 
overestimates $z$ and $z^\prime$, but underestimates 
$z^{\prime\prime}$, which results in a somewhat too high efficiency of 
the thermal diffusion (see~\much).

\subsection{Effect on solar models} 

\begin{figure}[t] 
\includegraphics[width=\columnwidth]{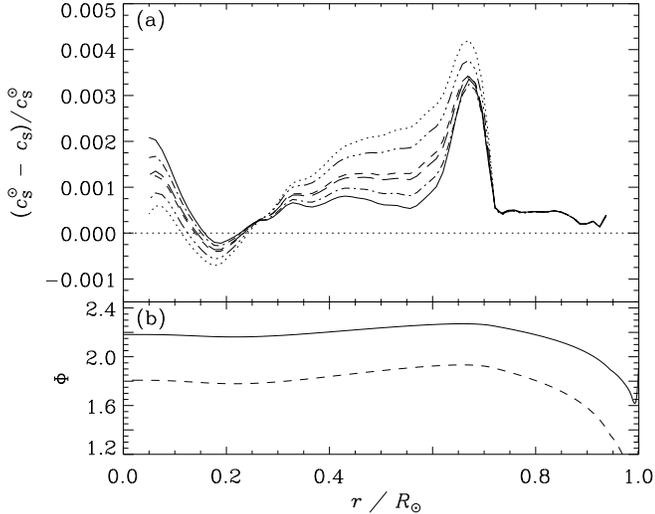} 
\caption{\textbf{a)}~The relative difference between the seismic model of 
\cite{BCC97} and solar models computed with different resistance 
coefficients: S1 (solid line), S2 (dotted), 
S3 (dash-dot-dot-dotted), S4 (short-dashed), S5 (long-dashed) and S6 
(dash-dotted). The resistance coefficients used in the models are 
summarized in Table~\ref{tab:phys}. \textbf{b)} The run of $\Phi$ 
for H$^+$-H$^+$ (solid) and He$^{2+}$-He$^{2+}$ collisions (dashed line) in solar 
models.\label{diffvgl}}    
\end{figure} 

\mybf{In order to estimate the influence of different choices of the
resistance coefficients on the solar structure, we computed various
solar models}
utilizing the
stellar evolution code and element diffusion routine described in 
\cite{S02} and references 
therein. Element diffusion is treated following the scheme by  
\thoul, considering in addition the effect of electron degeneracy and  
partial ionization of elements; the latter is implemented 
by defining a mean charged ion per element, instead of computing 
diffusion for each ion separately. This treatment of partial 
ionization is sufficient for a large range of stellar masses, including 
the Sun. \mybf{We neglect radiative levitation, which leads to a
tiny improvement of the theoretical 
sound-speed profile compared to the Sun of at most 0.025\% at $\mymbf{r\approx
0.6\,R_\odot}$ \citep[see][]{TRM98}}.

\begin{figure}[t] 
\includegraphics[width=\columnwidth]{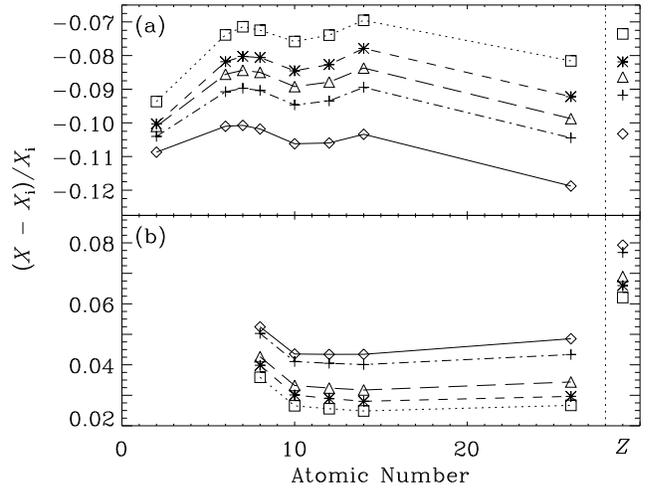} 
\caption{\textbf{a)} The surface depletion of various elements 
in solar models with different collision integrals from the pre-main 
sequence until now. The line-styles correspond to Fig.~\ref{diffvgl}. On 
the right hand side the change in the overall metallicity is 
shown. \textbf{b)} Like a), but for the centre.\label{abund}}    
\end{figure}

The sound speeds of various solar models computed using different 
choices of the collision  
integrals are summarized in Fig.~\ref{diffvgl}a, while the variation  
of He and metal abundances 
at the surface and at the centre of the Sun 
are shown in Fig.~\ref{abund}. The 
solid line represents a solar model (S1, see Table~\ref{tab:phys}) 
computed with $K$ of \iben\ and 
constant values for the $z$'s of \burg\, 
corresponding to the values used by \thoul. 
This description has been used, for instance, in  
solar models of \cite{Bah98} and \cite{S02}.  
Dropping the assumption of constant  $z$'s, and using instead the 
functions of \much, the sound-speed difference to a seismic model 
becomes about 4 times higher for $0.3 <r/R_\odot < 0.65$ (dotted line 
in Fig.~\ref{diffvgl}a; S2), caused by the reduced thermal diffusion 
efficiency. The latter leads also to a diminished surface depletion of He 
and metals (see Fig.~\ref{abund}). 
When one employs, in addition, \much's values for $K$ instead of the 
\iben\ ones, the difference to model S1 (dash-dot-dot-dotted 
line) lowers, because \much\ computed an about 7\% smaller value for $K$ than 
\iben\ (Fig.~\ref{fricvgl}a) for the relevant $\Phi$ range, which results
in slightly higher diffusion velocities.  

In solar conditions, \much's value of 2.62 for $z^{\prime\prime}$ is 
however too high with respect to the exact result; in fact 
$\Phi$ is about 2.2 in solar radiative regions (see Fig.~\ref{diffvgl}b), 
and $z^{\prime\prime}$ has to be between 2.3 and 2.4 (Fig.~\ref{fricvgl}d). 
Applying the correct $z^{\prime\prime}(\Phi)$ would further 
reduce the sound-speed difference compared to models computed with the 
value of \much.  
Indeed, using our formulae for \paq's collision 
integrals, which gives similar values of $K$ as \much\ 
(Fig.~\ref{fitvgl}a), but  includes the variation of 
$z^{\prime\prime}$, lowers the sound-speed difference 
(compare S3 and S4 in Fig.~\ref{diffvgl}a). 
A small additional increase of the diffusion velocities is obtained by 
considering different collision integrals for $e$-ion (attractive) and 
ion-ion or $e$-$e$ (repulsive) interactions (see also 
Fig.~\ref{abund}), which leads to model S5  
(long-dashed line in Fig.~\ref{diffvgl}a).

Typical additional quantities inferred by helioseismological methods are the 
depth of the convective zone and the surface helium abundance, which 
are determined to be $0.713\pm0.001$~\citep{BA97} and 
$0.246\pm0.002$~\citep{Sar}, respectively.  
In all models of Table~\ref{tab:phys} apart from S2 these quantities are well  
within the observational limits. Nonetheless, a general trend 
following the overall  
sound speed behaviour can be observed: The larger the sound-speed 
difference of the respective model the shallower the convective envelope 
and the higher the surface helium content.  

\begin{table}[t] 
\caption{Source for resistance coefficients used 
in the solar models discussed in the text and typical helioseismological quantities 
of the models. The column denoted 
with $z$ represents here $z$, $z^\prime$ and $z^{\prime\prime}$; 
column ``r/a'' shows models where repulsive and attractive potentials 
are distinguished, and ``qu.'' where quantum effects are 
included. Note, that in case of \paq\ our fitting formulae for the  
collision integrals have been used. The two columns on the r.h.s.\ denote 
the resulted depth of the convective envelope ($R_\mathrm{b.c.}/R_\odot$) and 
the surface helium abundance ($Y_\mathrm{s}$) in the respective solar 
models. \label{tab:phys}}   
\begin{tabular}{ccccccc} \hline\hline 
    & $K$   & $z$ &  r/a & qu. & $R_\mathrm{b.c.}/R_\odot$ & 
$Y_\mathrm{s}$ \\ \hline 
S1  & \iben & \burg               & --- & --- & 0.7125 & 0.2448 \\ 
S2  & \iben & \much               & --- & --- & 0.7151 & 0.2476 \\ 
S3  & \much & \much               & --- & --- & 0.7144 & 0.2467\\ 
S4  & \paq  & \paq                & --- & --- & 0.7134 & 0.2455\\ 
S5  & \paq  & \paq                &  X  & --- & 0.7135 & 0.2457 \\ 
S6  & \paq  & \paq                &  X  & X & 0.7128 & 0.2452 \\\hline   
\end{tabular} 
\end{table} 
 
\section{Quantum corrections} 
 
All computations of the 
resistance coefficient discussed in the previous section have assumed  
that the collisions are dominated by the classical interaction 
between two point-charge particles.  
For long-range potentials like the Debye-H\"uckel one, quantum effects 
become important at high energies, whereas the behaviour is classical 
at low energies. 
%In high density plasmas the classical treatment of particle collisions 
%becomes inaccurate.  
\citet[\hahn]{HMS71} have computed corrections 
to the classical collision integrals, such that  
\begin{eqnarray} 
{\Omega^{(p,q)}}^\ast & = &{\Omega^{(p,q)}_\mathrm{class}}^\ast(T^\ast) - 
\nonumber\\ 
&& \left({\Omega^{(p,q)}_\mathrm{qu}}^\ast(T^\ast{\Lambda^\ast}^2) \pm 
 \frac{1}{2s+1} 
 {\Omega^{(p,q)}_\mathrm{ex}}^\ast(T^\ast{\Lambda^\ast}^2) \right)\rm{,} \label{qudef} 
\end{eqnarray} 
where ${\Omega^{(p,q)}_\mathrm{class}}^\ast$ is the classical value 
of \S\ref{coll}, ${\Omega^{(p,q)}_\mathrm{qu}}^\ast$ contains the quantum  
mechanical correction, and ${\Omega^{(p,q)}_\mathrm{ex}}^\ast$ the 
exchange contribution, which has to be dropped for distinguishable 
particles (the indices $ij$ denoting different interacting 
particles have been omitted for the sake of clarity.) The upper sign in 
the term in parenthesis of Eq.~\bref{qudef} refers to particles  
obeying Fermi-Dirac statistic, the lower one for particles following 
Bose-Einstein statistics. The spin of the particle is denoted, as 
usual, by $s$, and $\Lambda^\ast$ is the de~Boer parameter 
given by 
\[ 
{\Lambda_{ij}^\ast}^2 = \frac{h^2}{2\mu_{ij} Z_i Z_j e^2 \lambda_\mathrm{D}}. 
\] 
Tabulated values for ${T^\ast}^2{\Omega^{(p,q)}_\mathrm{qu,ex}}^\ast$ 
are given in \hahn, which we again fitted by polynomials of 
5$^\mathrm{th}$ order (see Appendix~A).  
 
\begin{figure}[t] 
\includegraphics[width=\columnwidth]{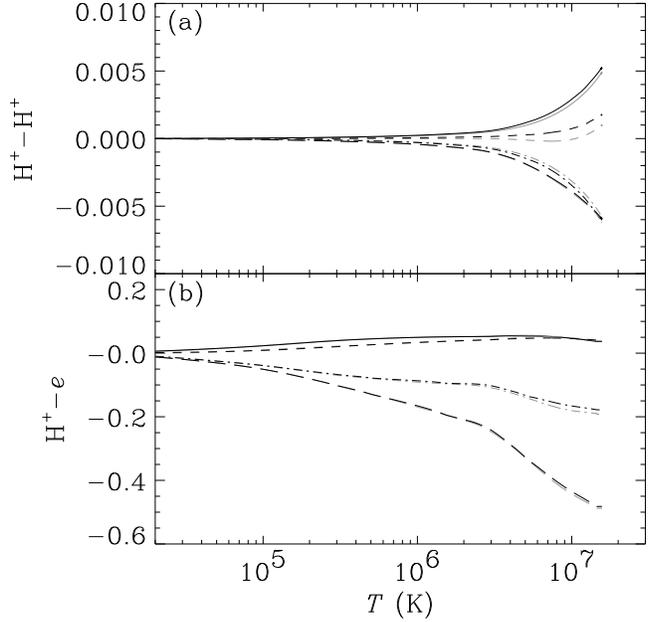} 
\caption{\textbf{a)}~The relative change of $K$ (long-dashed line), $z$ 
(solid), $z^\prime$ (short-dashed) and $z^{\prime\prime}$ 
(dash-dotted) for $\mathrm{H}^+$-$\mathrm{H}^+$ collisions in the Sun 
in the sense (quantum - classical)/classical.  
For the dark line the quantum corrections tabulated by \hahn\ have been 
utilized, while the grey lines were obtained with our fitting 
formulae (see Appendix~B). \textbf{b)}~As a), but for $\mathrm{H}^+$-$e$ 
collisions.\label{quvgl}}     
\end{figure} 
 
The excellent accuracy of our fitting formulae can be checked in 
Fig.~\ref{quvgl}, where the effect of the quantum corrections on the 
resistance coefficient and the thermal diffusion coefficients in the 
Sun is shown. The corrections increase toward the centre, and especially 
for the $\mathrm{H}^+$-$e$ collisions they considerably alter 
$K$. However, the diffusion velocity of H is basically determined by 
$\mathrm{H}^+$-$\mathrm{H}^+$ interactions and in that case the 
corrections for all quantities are below 1\%. The expected 
weak influence on the solar structure is demonstrated by model S6, 
where our fitting formulae to the quantum corrections of \hahn\ have 
been used. Overall, the diffusion velocities for all elements have 
been increased, more prominently in the centre, where the difference to 
models S5 (without quantum corrections) is the biggest 
(Fig.~\ref{quvgl}). This caused a further reduction of the sound-speed 
difference of model S6 compared to S5 (Fig.~\ref{diffvgl}a).  
 
It should be mentioned that the spin-dependent term in 
Eq.~\bref{qudef} would demand to know exactly the number of atoms 
in each ionization and in each excited state. However, the diffusion 
velocity of all elements is basically determined by their collision 
frequency with the most abundant element (H), and thus only for 
$\mathrm{H}^+$-$\mathrm{H}^+$ collision this spin-dependent term 
has to be accounted for (the spin-dependent term appears only in case of 
indistinguishable particles).  
Moreover, when quantum corrections become important inside stars most 
of the elements are already fully ionized. Thus only very tiny errors 
are introduced when, as in this work, only one mean-charged ion per 
element is considered. 
 
It is interesting to notice that model S6, which includes the most accurate collision 
integrals of \paq\, accounts for the difference between repulsive and 
attractive interactions, and includes quantum effects, has almost the 
same sound-speed profile as model S1 (cf.~Fig.~\ref{diffvgl}a), which contains the most crude 
treatment of diffusion among all models considered in this work. Also  
the depth of the convective envelope and the surface helium content 
are nearly identical (Table~\ref{tab:phys}). Therefore,  
just by chance, 
a very accurate solar model with respect to diffusion is obtained when 
using the resistance coefficients chosen by \thoul.

\section{Conclusions} 
 
In this paper, we have compared resistance
coefficients computed by various groups for the calculation of atomic
diffusion constants, and we have discussed the inclusion of quantum
effects on the otherwise classical determination of these quantities.
We provide simple analytical formulae to implement easily the
presently most accurate determination of resistance 
coefficients, i.e., the classical result by \paq\ plus the effect of 
quantum corrections determined by \hahn. (FORTRAN routines for
computing updated resistance coefficient and the consequent diffusion
constants are publicly available.\footnotemark[2])
 
The different classical computations based on a 
a Debye-H\"uckel type of potential produce solar sound-speed profiles
with significant differences when compared to the current accuracy of 
helioseismological determinations.
Just by chance, our accurate treatment including quantum corrections
produces a solar sound-speed profile comparable to the one obtained using
the less accurate  $z$, $z^\prime$  and 
$z^{\prime\prime}$ coefficients selected by \thoul.
\mybf{In this context, we would like to add that
with none of the descriptions for the collision integrals -- neither with the
inclusion of radiative levitation -- the 
bump in the sound-speed difference at $\mymbf{r\approx0.65\,R_\odot}$
disappears. So, an additional mixing process beyond the formal boundary
of the convective zones is still a probable candidate to resolve this
discrepancy \citep[see, e.g.,][]{Richard96}.}

The introduction of quantum corrections increases 
the efficiency of diffusion with respect to the classical case, and
their effect is more pronounced for higher densities.
It will therefore be important to test the effect of our accurate
resistance coefficients on models of objects denser than the Sun, like
white dwarfs, 
and main sequence turn-off stars of galactic globular clusters, 
where also radiative levitation is
altering the surface abundance patterns \citep{RMR02}.

Although the accuracy of the diffusion constants could be improved
considerably when using the correct functions for the resistance
coefficients, there still remains the limitation that Burgers'
formalism is equivalent only to \citeposs{CC70} second
approximation. This causes an intrinsic uncertainty when using
Burgers' equation of the order of 10\% \citep{RD82}, which is
difficult to reduce further.

\acknowledgement{H.S.\ has been supported by a Marie Curie Fellowship of the   
European Community programme ``Human Potential'' under contract number   
\mbox{HPMF-CT-2000-00951}.}

\normalsize

\appendix 
\section{} 
 
In this appendix we provide analytic formulae 
for ${T^\ast}^2\,{\Omega^{(s,t)}}^\ast$ and those combinations of the 
indices $s$ and $t$ which are needed to compute the resistance 
coefficients $K$, $z$, $z^\prime$ and 
$z^{\prime\prime}$ using Eq.~\bref{kdef}--\bref{zppdef}. The 
polynomial functions for the classical case  
have been obtained by a least square-difference 
fitting of the results of \paq. 
 
The best fits for the classical collision integrals could be obtained with  
\begin{equation}\label{fitatt} 
{T^\ast}^2\,{\Omega^{(s,t)}}^\ast = \sum_{k=0}^5 c^\mathrm{A}_{st}(k)\, 
\Phi^k  
\end{equation} 
for attractive and  
\begin{eqnarray} 
{T^\ast}^2\,{\Omega^{(s,t)}}^\ast & = &\sum_{k=0}^3 c^\mathrm{R}_{st}(k)\, 
(\tilde{\Phi}_{st})^k \quad\mbox{with} \\ 
\tilde{\Phi}_{st} & = & \ln\left(\ln(1 + 
\Lambda^{b_{st}})\right)\label{nphidef}  
\end{eqnarray} 
for repulsive potentials. We dropped the indices $i, j$ for the sake 
of clarity. 
The numerical values for the coefficients 
$c^\mathrm{A}$ and $c^\mathrm{R}$, and the exponents $b$ are 
summarized in Table~\ref{tab:fit}. Notice that  
it is sufficient to determine  
${T^\ast}^2\,{\Omega^{(s,t)}}^\ast$, as in the definitions of the 
resistance coefficients the factor ${T^\ast}^2$ always cancels out. 
%In order to display our fits to \paq\ results for the repulsive 
%potential in Fig.~\ref{fitvgl}, we have computed the  
%collision integrals and resistance coefficients using  
%$\tilde{\Phi}_{st}$ has independent variable, and then transformed  
%$\tilde{\Phi}_{st}$ into $\Phi$. 
 
\begin{table} 
\caption{Coefficients for 
Eqs.~\bref{fitatt}--\bref{nphidef}.\label{tab:fit}} 
\setlength{\tabcolsep}{.56em} 
\begin{tabular}{lcccc}\hline\hline 
                  & \multicolumn{4}{c}{(s,t)} \\ 
                  & (1,1) & (1,2) & (1,3) & (2,2) \\ \hline 
$c^\mathrm{A}$(0) & $-$1.577 & $-$2.062 & $-$2.472 & $-$1.776 \\ 
$c^\mathrm{A}$(1) & 0.6285  & 0.5066 & 0.4452 & 0.8555 \\ 
$c^\mathrm{A}$(2) & 0.08141 & 0.05224 & 0.04911 & 0.07976 \\ 
$c^\mathrm{A}$(3) & 0.03769 & 0.03302 & 0.02851 & 0.01198 \\ 
$c^\mathrm{A}$(4) & $-$0.002702 & $-$0.0005197 & 
                  $1.605\!\times\!10^{-5}$ & $-$0.002642 \\ 
$c^\mathrm{A}$(5) & $-$0.001587 & $-$0.001291 & $-$0.001014 & 
                  $-$0.0004393 \\ \hline 
$c^\mathrm{R}$(0) & $-$1.862 & $-$2.465 & $-$2.857 & $-$1.702 \\ 
$c^\mathrm{R}$(1) & 2.313  & 1.667 & 1.386 & 1.916 \\ 
$c^\mathrm{R}$(2) & $-$0.1550 & $-$0.07154 & $-$0.05136 & $-$0.08114 
                  \\ 
$c^\mathrm{R}$(3) & $-$0.07188 & $-$0.03209 & $-$0.01620 & $-$0.04868 
                  \\ \hline 
$b$ & 0.6339 & 0.8228 & 0.9231 & 0.7807 \\ \hline 
\end{tabular} 
\end{table} 
 
The quantum-mechanical (``qu'') and exchange contribution (``ex'') can 
both be computed  
according to the following relationship 
\begin{eqnarray} 
{T^\ast}^2{\Omega^{(p,q)}_\mathrm{qu,ex}}^\ast 
(T^\ast\!{\Lambda^\ast}^2)& = &\frac{p 
N_p}{q(q+1)}\, 
I^{(q)}_\mathrm{qu,ex}(T^\ast\!{\Lambda^\ast}^2)\quad\mbox{with}\\ 
N_p^{-1} & = & 1 - \frac{1+(-1)^p}{2(1+p)}. 
\end{eqnarray} 
 
In these formulae the quantum and exchange parts of the collision 
integrals are expressed as functions of the quantity
$T^\ast\!{\Lambda^\ast}^2$.  
The best fits to the values tabulated in \hahn\ have been obtained 
with 
\begin{eqnarray} 
I^{(q)}_\mathrm{C}(x) & = & \sum_{i=0}^4 d_\mathrm{C}^{(q)}(i) 
\,{\xi_\mathrm{C}^{(q)}}^{i}(x)\mbox{,\quad where} \label{idef}\\ 
\xi_\mathrm{C}^{(q)}(x) & = & \left\{ 
\begin{array}{lcl} 
\ln(1 + g_\mathrm{C}^{(q)}\!\times\! x ) & \mbox{for}& 
\mathrm{C}=\mbox{``qu''} {\rm,}\\  
\ln(1 + g_\mathrm{C}^{(q)}\;/\; x )& \mbox{for}& 
\mathrm{C}=\mbox{``ex''} {\rm}\\ 
\end{array} 
\right.\label{xidef} 
\end{eqnarray} 
with $x=T^\ast\!{\Lambda^\ast}^2$ and
the additional constraint that $\xi^{(q)}_\mathrm{ex}(x) = 0$ for 
$x \le 0.1$. The index ``C'' denotes the appropriate 
expression for either the quantum 
or the exchange contribution to the collision integrals. 
The respective coefficients are provided in Table~\ref{tab:quf}. 
Note that $I^{(1)}_\mathrm{ex}=0$ and that Eqs.~\bref{idef} and 
\bref{xidef} are only valid for $x<1\,000$, as no values were provided 
by \hahn\ for larger $x$. We suggest not to extrapolate 
$T^\ast\!{\Lambda^\ast}^2$ beyond this upper boundary, but to use the 
values at  $T^\ast\!{\Lambda^\ast}^2=1\,000$, which are 
$I^{(2)}_\mathrm{ex} = 0.4913$, $I^{(3)}_\mathrm{ex} = 0.4960$,  
$I^{(1)}_\mathrm{qu} = 1.5053$, $I^{(2)}_\mathrm{qu} = 1.9581$ and  
$I^{(3)}_\mathrm{qu} = 2.2010$. 
 
\begin{table} 
\caption{Coefficients for 
Eqs.~\bref{idef}--\bref{xidef}\label{tab:quf}}  
\setlength{\tabcolsep}{1em} 
\begin{tabular}{lccc} \hline \hline 
                 & \multicolumn{3}{c}{$q$}  \\ 
                 &    1    &   2     &   3 \\ \hline     
$d_\mathrm{ex}(0)$ & --- & $-$0.7074 & $-$0.7103  \\ 
$d_\mathrm{ex}(1)$ & --- & $-$1.985 & $-$2.744 \\ 
$d_\mathrm{ex}(2)$ & --- & 0.5233 & 0.1870 \\ 
$d_\mathrm{ex}(3)$ & --- & $-$0.3270 & 0.03388 \\ 
$d_\mathrm{ex}(4)$ & --- & 0.03920 & $-$0.06502 \\ \hline 
$d_\mathrm{qu}(0)$ & 0 & 0 & 0 \\ 
$d_\mathrm{qu}(1)$ & 0.05333 & 0.03051 & 0.3133 \\ 
$d_\mathrm{qu}(2)$ & 0.01889 & 0.01021 & 0.03982 \\ 
$d_\mathrm{qu}(3)$ & 0.007114 & 0.009180 & $-$0.003032 \\ 
$d_\mathrm{qu}(4)$ & $-$0.0006953 & $-$0.0007776 & 
$1.388\!\times\!10^{-5}$ \\ \hline  
$g_\mathrm{ex}$ & ---  & 6.121 & 2.285 \\ 
$g_\mathrm{qu}$ & 0.3 & 1.0 & 0.15 \\ \hline 
\end{tabular} 
\end{table}

\section{} 
Here we describe the modifications to be applied to the 
routine by \thoul, 
in order to use any preferred formulation of the resistance 
coefficients.  
 
The first change involves Eqs.~(9)--(10) of \thoul, 
where the \iben\ definition of $K$ is implemented. They have to be 
replaced with the chosen representation of $K$. 
 
The matrix of elements $Y_{st}$, introduced at page 830 
of \thoul, has to be changed to 
\begin{eqnarray} 
Y_{st}= 3 y_{st}+ z^\prime_{st}\, x_{st} \, 
m_{t}/m_{s}. 
\end{eqnarray} 
 
Finally, the following elements of the matrix $\Delta_{ij}$ 
as defined in Eqs.~(33) and (34) of \thoul\ become
\begin{equation} 
\Delta_{ij} =  \left\{ 
\begin{array}{lcl} 
\sum_{k\neq i} z_{ik} k_{ik} x_{ik} & \mbox{for} & j=i+S{\rm,}\\*[6pt] 
-z_{iq}\, k_{iq}\, y_{iq} & \mbox{for} & j=S+1,\dots,2S \\ 
&& \mbox{and~}j\ne i+S{\rm,}\\ 
\end{array}  
\right. 
\end{equation} 
for $i=1,\dots,S$ and  
\begin{eqnarray} 
\Delta_{ij} =  \left\{ 
\begin{array}{lcl} 
2.5 \sum_{k\neq j} z_{pk} k_{pk} x_{pk} & \mbox{for} & j=p{\rm,}\\*[6pt] 
-2.5 z_{pj} k_{pj} x_{pj} & \mbox{for} & 
j=1,\dots,S\mbox{~and~}j\ne p{\rm,}\\*[6pt] 
\lefteqn{-{\textstyle\sum_{k\neq i}}k_{pk} y_{pk}(0.8z^{\prime\prime}_{pk} 
x_{pk}+Y_{pk})-0.4z^{\prime\prime}_{pp} k_{pp}} \\ 
& \mbox{for} & j=i{\rm,}\\*[6pt] 
\lefteqn{k_{pq} y_{pq} x_{pq} 
(3 + z^{\prime}_{pq} - 0.8z^{\prime\prime}_{pq})} \\ 
& \mbox{for} & j=S+1,\dots,2S\\ 
&&	\mbox{and~}j\ne i{\rm,} 
\end{array} 
\right. 
\end{eqnarray} 
for $i=S+1,\dots,2S$, where $p=i-S$ and $q=j-S$. Here, the same 
symbols have been used as in \thoul, to which the reader is referred for 
their definitions. 

%\bibliographystyle{aa} 
%\bibliography{mylit,stell,sun,helios} 

\newcommand{\singlet}[1]{#1}

\end{document}